\documentclass[aps,prl,twocolumn,superscriptaddress,showpacs]{revtex4-1}

\usepackage{graphicx}
\usepackage{subfigure}
\usepackage{dcolumn}
\usepackage{bm}
\usepackage{amsmath,amssymb}
\usepackage{ulem}
\usepackage[
  colorlinks=true,
  allcolors=blue,
  breaklinks=true
]{hyperref}
\makeatletter
\def\captionof#1#2{{\def\@captype{#1}#2}}
\makeatother

\usepackage[english]{babel}
\usepackage{times}
\usepackage{amsfonts}
\usepackage{psfrag}
\usepackage{verbatim}
\usepackage{color}

\begin{document}


\title{Phase co-existence in bidimensional passive and active dumbbell systems}

\author{Leticia F. Cugliandolo} 
\affiliation{Sorbonne Universit\'es, Universit\'e Pierre et Marie Curie - Paris VI,
Laboratoire de Physique Th\'eorique et Hautes \'Energies,
4 Place Jussieu, 75252 Paris Cedex 05, France}
\affiliation{Kavli Institute for Theoretical Physics, University of California at Santa Barbara, CA 93106, USA}

\author{Pasquale Digregorio}
\affiliation{Dipartimento di Fisica, Universit\`a degli Studi di Bari and
INFN, Sezione di Bari, via Amendola 173, Bari, I-70126, Italy}

\author{Giuseppe Gonnella}
\affiliation{Dipartimento di Fisica, Universit\`a degli Studi di Bari and
INFN, Sezione di Bari, via Amendola 173, Bari, I-70126, Italy}

\author{Antonio Suma} 
\affiliation{SISSA - Scuola Internazionale Superiore di Studi Avanzati,
Via Bonomea 265, 34136 Trieste Italy}

\received{\today}


\begin{abstract}
We demonstrate that there is macroscopic co-existence 
between regions with hexatic order and regions in the liquid/gas phase over a finite 
interval of packing fractions in active dumbbell systems  with repulsive power-law interactions
in two dimensions. 
In the passive limit this interval remains finite, similarly to what has been \textcolor{black}{found} in 
bidimensional systems of hard and soft disks. \textcolor{black}{We did not find discontinuous behaviour upon increasing activity from the passive limit.} 
\end{abstract}


\maketitle 

\setlength{\textfloatsep}{10pt} 
\setlength{\intextsep}{10pt}

Interest in the behaviour of  $2d$ (and also $3d$) macroscopic systems under continuous and homogeneous input of energy has been boosted by
their connection with active matter~\cite{Toner05,Fletcher09,Romanczuk12,Vicsek12,Cates12,Marchetti13,Marenduzzo14,Elgeti15}. 
This new type of matter can be realised in various  ways. Systems of self-propelled particles 
constitute an important subclass, with natural examples such as suspensions of bacteria~\cite{wu,Dombrowski04,rabani-ariel}, and artificial ones
made of Janus~\cite{Paxton04,Buttinoni13b,PhysRevX.5.011004} or asymmetric granular~\cite{Dauchot15} particles. In all these cases, the constituents consume  internal or environmental energy and use it to displace.  Very rich collective motion  arises 
under these out of equilibrium conditions, and liquid, solid and {\it segregated} phases
are observed~\cite{Tailleur08,Fily12,Stenhammer13,Redner13,Winkler14,weber2014defect}.
\textcolor{black}{In particular, in active Brownian particle systems, 
segregation, also called motility induced phase separation (MIPS), was claimed to occur
 {\it only above} a large critical threshold of the activity~\cite{Redner13,stenhammar2014phase,Suma14,marchetti2016minimal,redner2016classical}.} 

\textcolor{black}{Besides, the behaviour of passive disks is a classic theme of study in soft condensed matter.}
Recently, Bernard \& Krauth argued that $2d$ melting of hard and soft repulsive disks 
occurs in two steps, with a continuous Berezinskii-Kosterlitz-Thouless transition between the solid and 
\textcolor{black}{hexatic} phases, and a first order transition between the \textcolor{black}{hexatic}  and liquid 
phases, when density or packing fraction are decreased at constant temperature~\cite{BeKr11}.
The \textcolor{black}{hexatic} phase has  no positional order but quasi long-range orientational order, 
while the solid phase has quasi long-range positional  and proper long-range orientational
order. Liquid and quasi long-range orientationally ordered zones co-exist close to the liquid phase, within a narrow interval 
of packing fractions.

In this Letter we study  the phase diagram of a bidimensional model of active purely repulsive dumbbells 
and show that it does not comply \textcolor{black}{with the MIPS scenario}. 
We prove that the phase separation found at high values of the activity continuously links, in the passive limit, to a finite co-existence region
as the one predicted by Bernard \& Krauth for $2d$ melting of hard and soft repulsive disks~\cite{BeKr11}. There is no
non-vanishing critical value of activity needed for segregation in this system, making the popular MIPS scenario at least not general.

\begin{figure}[t]
\includegraphics[width=1.0\columnwidth]{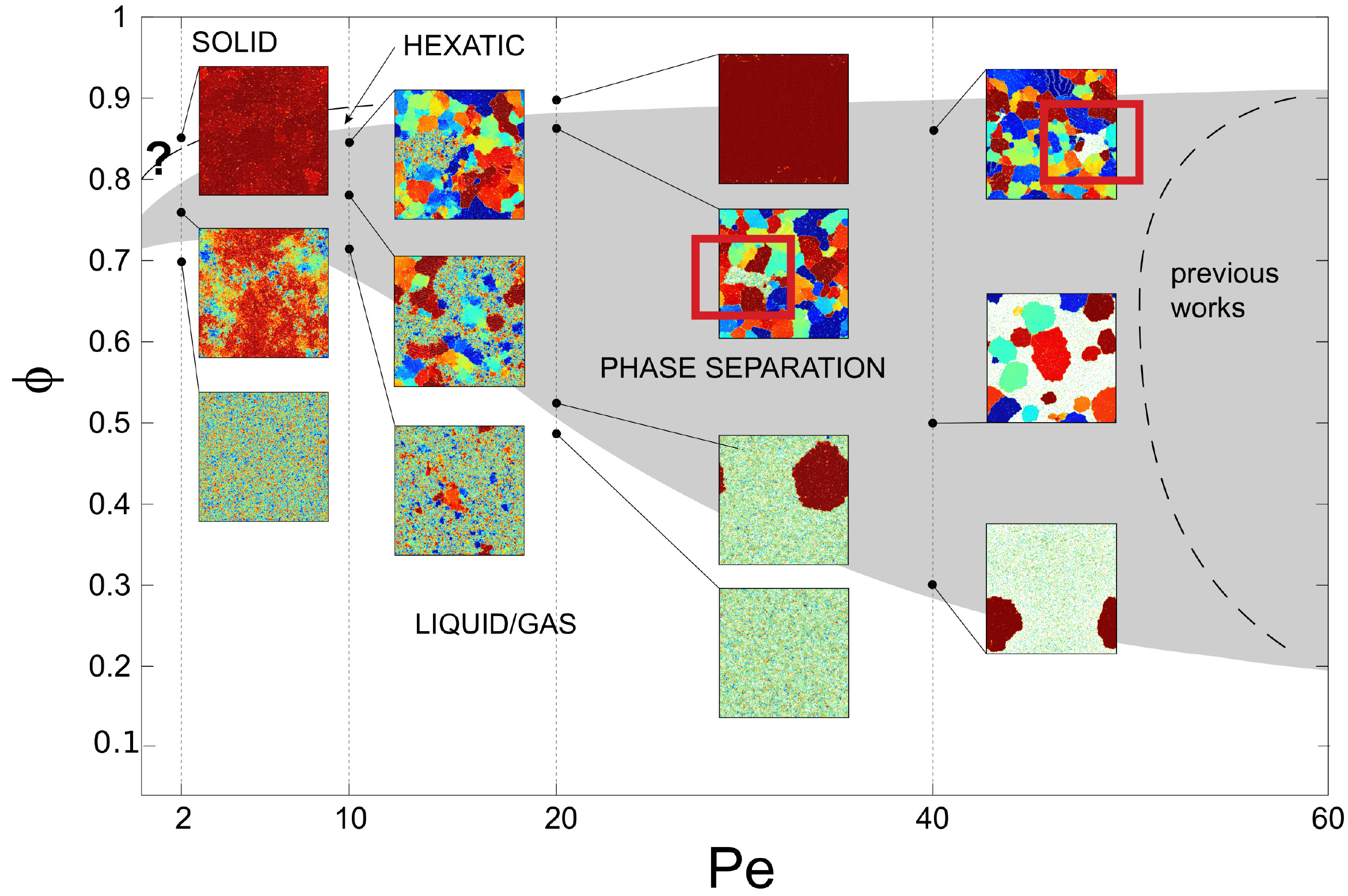}
\caption{(Color online.) The phase diagram and some representative local \textcolor{black}{hexatic} parameter maps. Note the red rectangular contours in the boxes
at Pe $= 20$ and Pe $= 40$ that surround the disordered regions. 
The way in which the phase boundaries are determined is explained in the text and more details are given in the SM.}
\label{fig:phase-diagram}
\end{figure}

The reason for choosing a dumbbell model is that 
many natural swimmers have elongated shape, and a hard dimer is the simplest approximation of such anisotropy~\cite{Frenkel91,Frenkel93,valeriani2011colloids}.
This geometry  favours aggregation at intermediate  
densities and sufficiently strong activation~\cite{Peruani06,Suma13,Suma14,Suma14c,Joyeux16,gonn15,siebert2016phase}.
In this limit the evolution of an initial homogeneous phase occurs 
by nucleation and growth of clusters~\cite{Suma13} and the system  phase separates.
At the other extreme, for sufficiently low  densities and not so strong activity, particles form only very small 
clusters that  do not coalesce~\cite{Suma14b,Suma14c,Suma15a}.
The results in this Letter complement these two extreme limits. In the absence of activity we confirm the results  of 
Bernard, Kapfer \& Krauth for hard and soft disks~\cite{BeKr11,Engel13,KaKr15} --for what concerns the existence of a first-order transition from the liquid phase-- 
using now a molecular system and estimating the density interval for co-existence.
We prove that this interval {\it continuously expands} towards the strong activity region where 
cluster aggregation had already been observed. Hence, there is no discontinuity between the 
passive and active regions in the phase diagram with phase separation. Figure~\ref{fig:phase-diagram}
summarises this scenario that, we emphasise,  is different from what has been stated in the literature 
so far. \textcolor{black}{We did not analyze in this work the transition between hexatic and solid phase. }

Event-chain algorithms have proven to be an efficient tool to equilibrate
$2d$ interacting particle systems~\cite{BeKrWi09} and they have been used to give strong support to the 
two step phase transition scenario~\cite{BeKr11,KaKr15}.
We use, however, conventional molecular dynamics in order to simulate the
out of equilibrium dynamics of active  systems as realistically as possibly.  

The model consists of $N$ diatomic molecules (dimers)
with identical spherical head and tail centered at a fixed distance equal to their diameters, 
$\sigma_{\rm d}$.  
Interactions are mediated by a purely repulsive potential,
$U( r )=4\epsilon[(r/\sigma)^{-2n} - (r/\sigma)^{-n}]$,
truncated at its minimum $r_c=2^{1/n}\sigma$, with  $r$ the distance between the centers of any two 
disks. We set $\sigma=2^{-1/n}\sigma_{\rm d}$, so that $r_c=\sigma_{\rm d}$
and we favor co-existence  in the passive system~\cite{KaKr15}
using $n=32$, \textcolor{black}{with particle overlap unlikely smaller than $\sigma_{\rm d}$.}
\textcolor{black}{Results similar to those shown in the following have been obtained  for  $n=6$, but with a narrower 
coexistence region in the passive limit.}
The evolution of the position ${\mathbf {r}_i}$ of the $i$-th bead is given by a Langevin equation,
\begin{equation}
m_{\rm d} \ddot {\mathbf r}_i=-\gamma_{\rm d}\dot {\mathbf r}_i -{\boldsymbol \nabla}_i U+
{\boldsymbol F}_{\rm act}+
\sqrt{2k_BT\gamma_{\rm d}} \, {\boldsymbol \eta}_i(t) \; .
\label{bd}
\end{equation}
$\gamma_{\rm d}$ is the friction coefficient,
${\boldsymbol \nabla}_i=\partial_{\mathbf{r}_i}$,
$T$ is the temperature of the thermal bath,
$m_{\rm d}$  is the mass of a bead, ${\boldsymbol F}_{\rm act}$
is a tail-head-directed active force
with constant magnitude $F_{\rm act}$,
and ${\boldsymbol \eta}_i(t)$ is an uncorrelated Gaussian noise with zero 
mean and unit variance. 
We set the parameters to be in the over-damped limit~\footnote[1]{All physical quantities are expressed in reduced 
units~\cite{allen} of the sphere's mass $m_{\rm d}$,  diameter $\sigma_{\rm d}$ and potential energy $\epsilon$. 
The time unit is the standard Lennard-Jones time 
$\tau_{LJ}=\sigma_{\rm d} (m_{\rm d}/\epsilon)^{1/2}$.
Other important simulation parameters, in reduced units, are 
$\gamma_{\rm d}=10$, $k_BT=0.05$ and we set
$k_B=1$.}. The dimensionless control parameters are the area fraction covered by
the active particles, $\phi=N\pi\sigma_{\rm d}^2/(2A)$, where $A=L^2$ is the area 
of the simulation domain, and  the P\'eclet number, Pe = 2$F_{\rm act}\sigma_{\rm d}/(k_BT)$.
We used $L = 500\, \sigma$ and periodic boundary conditions. Each run took, 
typically, $5 \times 10^5$ simulation time units (MDs~\footnote[1] ~~not written henceforth). We performed tests in systems with 
$L \approx 1500 \, \sigma$ run for longer and we did not
find differences with the results shown. More details on the algorithm and running-times are given 
in the Supplemental Material (SM).

We quantify our assertions with the measurement of  the
{\it local densities} $\phi_j$ (computed in two ways explained in the SM), 
and the {\it local \textcolor{black}{hexatic} parameter} evaluated as
\begin{equation}
 \psi_{6j}=\frac{1}{N^j_{\rm nn}}\sum_{k=1}^{N^j_{\rm nn}}e^{6{\rm i}\theta_{jk}},
\end{equation}
where $N_{\rm nn}^j$ are the nearest neighbours 
of bead $j$ found with a Voronoi tessellation algorithm~\cite{voro++} and 
$\theta_{jk}$ is the angle between the segment that connects $j$ with its neighbour $k$ 
and the $x$ axis. For beads regularly 
placed on the vertices of a triangular lattice, each site has six nearest-neighbours, $\theta_{jk}=2k\pi/6$, and $\psi_{6j} =1$.
We also consider the modulus of the average per particle and the average per particle of the modulus, 
\begin{eqnarray}
2N \, \psi_6 \equiv  \Big{|} \sum_{j=1}^N \psi_{6j} \Big{|}
\; , 
\qquad
2N \, \Gamma_6 \equiv  \sum_{j=1}^N |\psi_{6j}|
\; .
\label{eq:defs}
\end{eqnarray}
We visualise the local values of $\psi_{6j}$ as proposed in~\cite{BeKr11}:
first, we project the complex local values $\psi_{6j}$ onto the direction of their space average, 
next, each bead is colored according to this normalised projection.
Zones with orientational order have uniform color, whatever it is.

\begin{figure}[t]
\begin{center}
\includegraphics[width=0.98 \columnwidth]{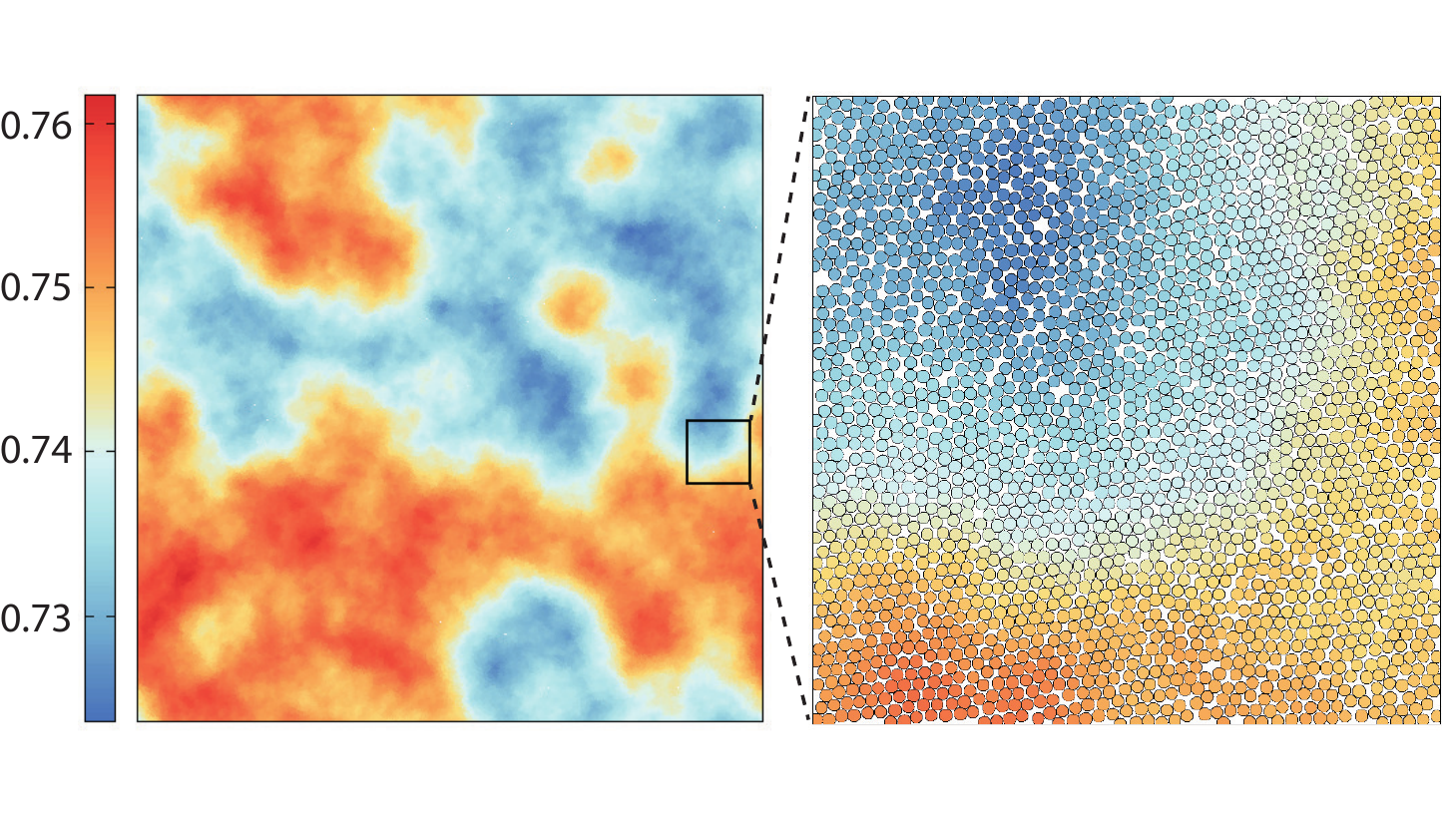}
\end{center}
\vspace{-0.65cm}
\caption{(Color online.)
Passive system with $\phi = 0.74$ in equilibrium. 
Left: the density plot constructed by averaging the local density
over a region with radius equal to $20 \, \sigma$ \textcolor{black}{(similar results are obtained averaging over a region with 
radius in $[10, 50]$)}. Right: a zoom over the region within the black square in the 
left plot showing the individual dumbbells 
close to the interface between a region with \textcolor{black}{hexatic} order (lower right) and 
a disordered sector (upper left). The color code is the same as in the left panel. The local \textcolor{black}{hexatic} map for this 
configuration is shown in the central upper panel in Fig.~\ref{fig:tres}.
}
\label {fig:dos}
\end{figure}

We start by studying the passive system. We use three kinds of initial states: random configurations with 
positions and orientations uniformly distributed, striped states with an ordered close-packed slab, and a \textcolor{black}{hexatic}-ordered state (see Sec.~S1 in the SM).
In all cases we present data evolved for sufficiently long 
to ensure that the initial state is forgotten and equilibration is reached. In the SM we exemplify  
the transient.

For $\phi < 0.730$ any initial state with phase separation 
quickly melts and eventually evolves as a liquid.
This is confirmed by the fact that translational and \textcolor{black}{hexatic} 
correlation functions decay exponentially with distance.
Above $\phi \approx 0.756$ initial states with \textcolor{black}{hexatic} order remain ordered and the
 correlations decay very slowly (see Fig.~S3). 
In between there is a regime with co-existence, as we now prove.

The first evidence for co-existence is given in Fig.~\ref{fig:dos} where we show the local density plot in equilibrium, with a zoom close to an interface between dense and sparse regions. 
The \textcolor{black}{hexatic} order in the region with high density and the lack of 
orientational order in the sparse region are clear in the zoom. 

\begin{figure}[t]
\begin{center}
\includegraphics[width=1.0 \columnwidth]{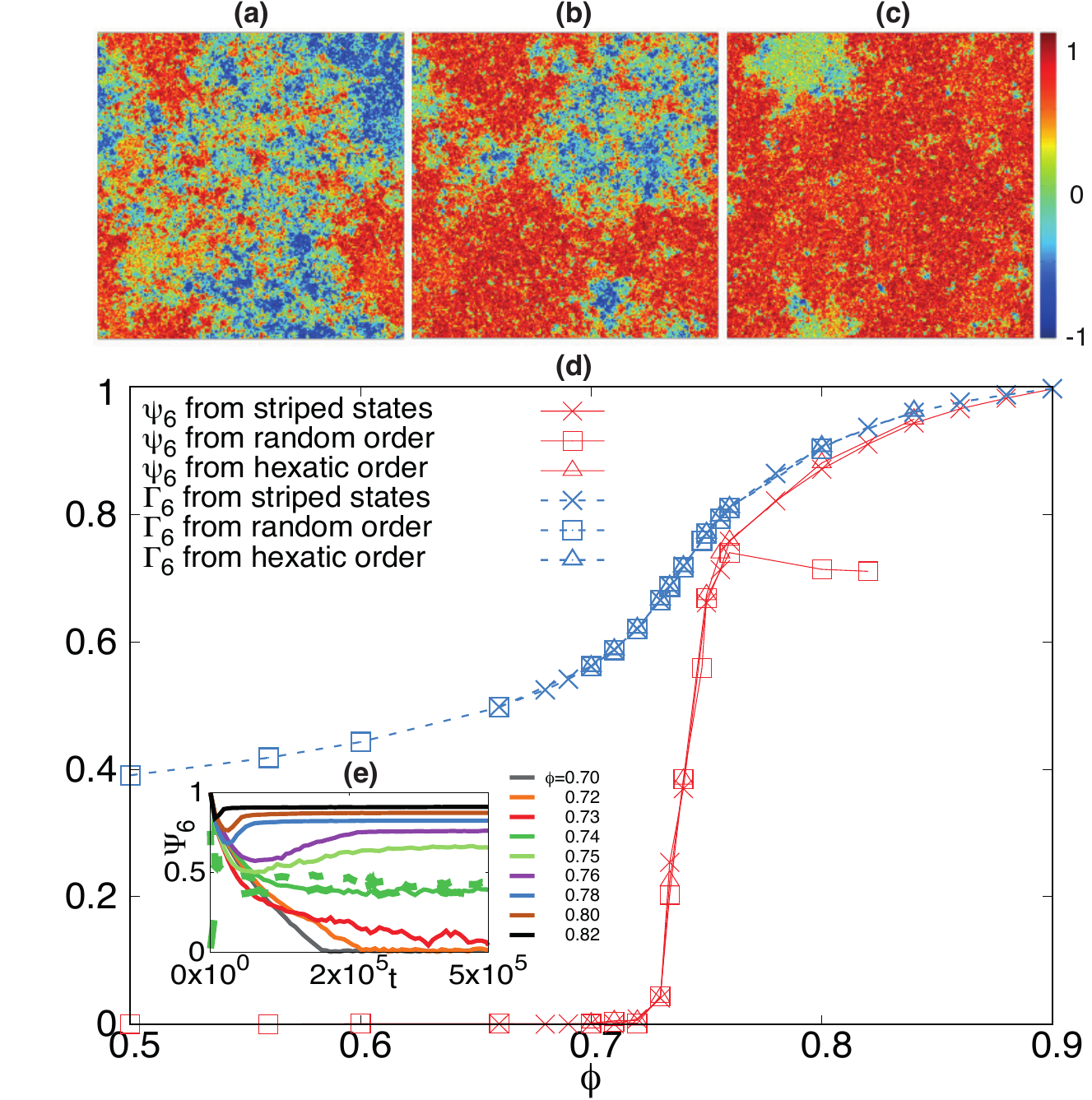}
\end{center}
\vspace{-0.65cm}
\caption{(Color online.)
Passive system.
\textcolor{black}{(a-c)} The local \textcolor{black}{hexatic} parameter, $\psi_{6j}$, in equilibrium configurations at $\phi=0.734~\textcolor{black}{\textrm{(a)}}, \ 0.740~\textcolor{black}{\textrm{(b)}}, \ 0.750~\textcolor{black}{\textrm{(c)}}$. 
\textcolor{black}{(d)} $\psi_{6}$ and  $\Gamma_{6}$
defined in Eqs.~(\ref{eq:defs}) as functions of $\phi$. 
\textcolor{black}{(e)} Time evolution of $\psi_6$ from striped initial conditions
at various densities with different solid lines. 
We also display the evolution at $\phi=0.74$ from random and \textcolor{black}{hexatic}-ordered initial states with dashed and dotted lines. 
}
\label {fig:tres}
\end{figure}

Further evidence for co-existence at this and other global densities
is given in \textcolor{black}{Fig.~\ref{fig:tres}(a-c)} where we show the local \textcolor{black}{hexatic} parameter on 
three equilibrium snapshots at $\phi=0.734, \ 0.740, \ 0.750$. 
These configurations are chosen at the long-time limit of the 
evolution of random initial states. 
The regions with local \textcolor{black}{hexatic} order are 
also regions of local high density and, conversely, in the sparse regions the 
dumbbells do not have orientational order (see Fig.~S4 in the SM where the corresponding 
density plots as the one in Fig.~\ref{fig:dos}
and histograms of the local densities are shown).
In \textcolor{black}{Fig.~\ref{fig:tres}(d)} we display the asymptotic $\psi_{6}$ and $\Gamma_6$ defined in Eq.~(\ref{eq:defs}) 
against $\phi$ for the three kinds
of initial conditions.  The data have been averaged 
over the last ten configurations (sampled every $\sim \, 10^4$).
The results confirm that the departing state is forgotten as the 
curves coincide within numerical accuracy. The $\psi_6$  curve for  
random initial configurations at $\phi\gtrsim0.780$  is an exception and it still has to undergo
a coarsening process to orient the clusters in the same direction, see \textcolor{black}{Fig.~S7} in the SM.
All curves increase with 
$\phi$ indicating that the proportion of regions with \textcolor{black}{hexatic} order with respect to the 
disordered ones grows with $\phi$. The curve $\Gamma_6$ against $\phi$ is continuous 
and smooth while the one for $\psi_6$ although also continuous, shows a very steep increase
starting at the smallest density at which 
co-existence appears. In \textcolor{black}{Fig.~\ref{fig:tres}(e)}  we show the time-dependence
of $\psi_6$. 
The asymptote vanishes for $\phi<0.730$ but grows with $\phi$ for $\phi>0.730$. 
For $\phi=0.740$ we follow  the evolution 
of different kinds of initial states to prove that they all approach the same asymptote. 
The evolution of the local $\psi_{6j}$ for these three initial conditions is illustrated in Fig.~S2.
The last one  is at a time at which the (green) curves in Fig.~\ref{fig:tres} have reached the 
plateau. Additional  signatures of liquid, coexisting and ordered phases are 
given in Fig.~S5 that shows the structure factor for six $\phi$s.

Turning these arguments into a quantitative analysis, we find co-existence in the passive system 
in the interval $\phi \in [0.730, \ 0.756]$, approximately, justifying the extent of the grey region
on the Pe = 0 axis in Fig.~\ref{fig:phase-diagram}.

\begin{figure}[b]
\begin{center}
\includegraphics[width=1.0 \columnwidth]{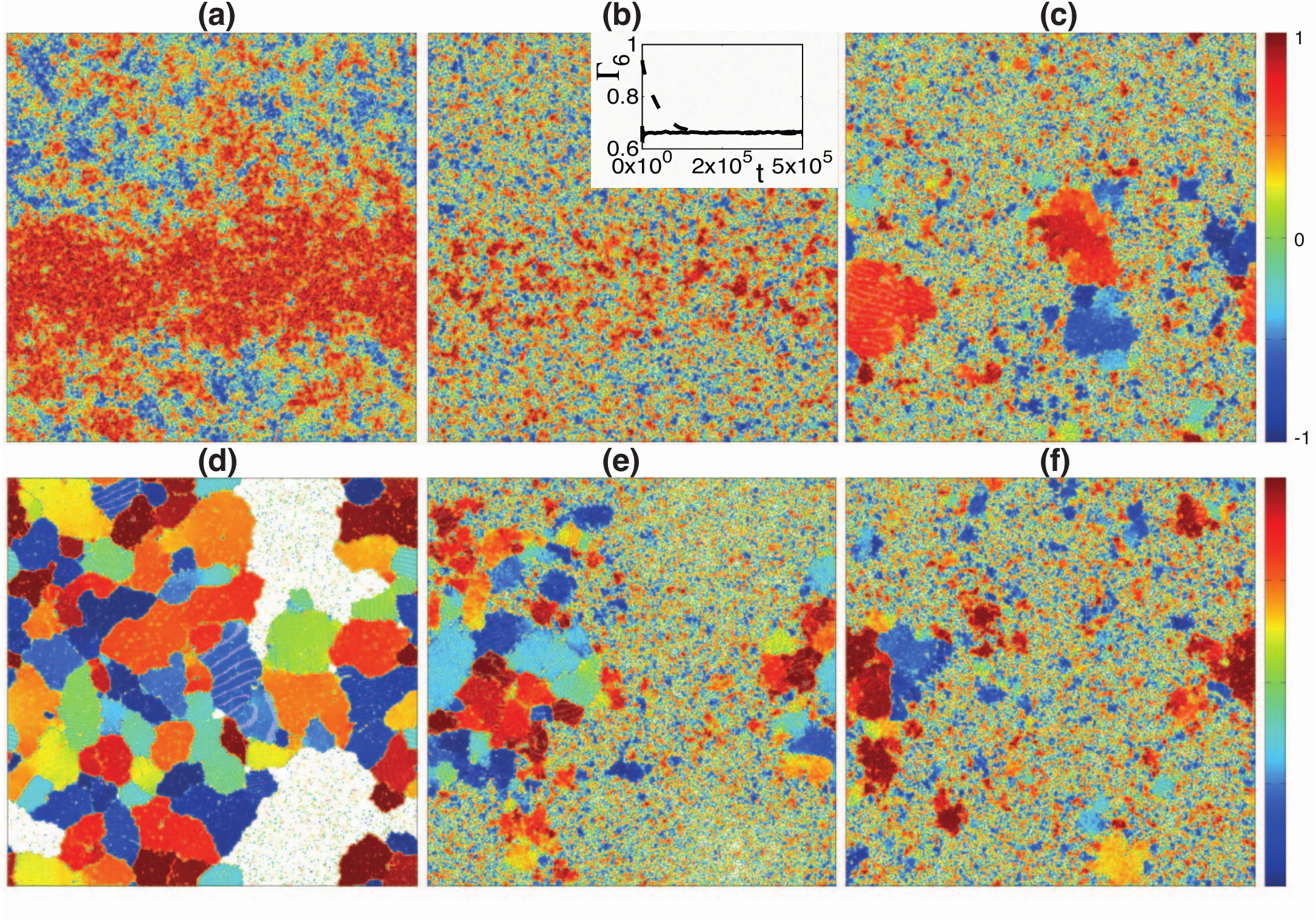}
\end{center}
\vspace{-0.4cm}
\caption{(Color online.)
Time evolution at Pe = 10 and $\phi=0.734$. 
The initial states for the sequences \textcolor{black}{(a-c)} and \textcolor{black}{(d-f)}
correspond to \textcolor{black}{(a)} a phase separated configuration at Pe = 0 evolved from a configuration with 
a striped-state of dumbbells
and \textcolor{black}{(d)} the steady state at Pe = 40. The system is in its 
gas phase in the white zones (cfr. Fig.~S12 (d) in the SM).
The times at which the subsequent snapshots are taken are 
$t  =  2.5 \times 10^2~\textrm{(b)}, \, 8.6 \times 10^5~\textrm{(c)}$  and $t \, = \, 10^5~\textrm{(e)}, \, 6 \times 10^5~\textrm{(f)}$. 
 As an inset in \textcolor{black}{(b)} we include the time-dependence of $\Gamma_6$ for the 
 two runs.
 }
\label{fig:cuatro}
\end{figure}

\begin{figure}[t]
	\centering
	\includegraphics[width=0.9\columnwidth]{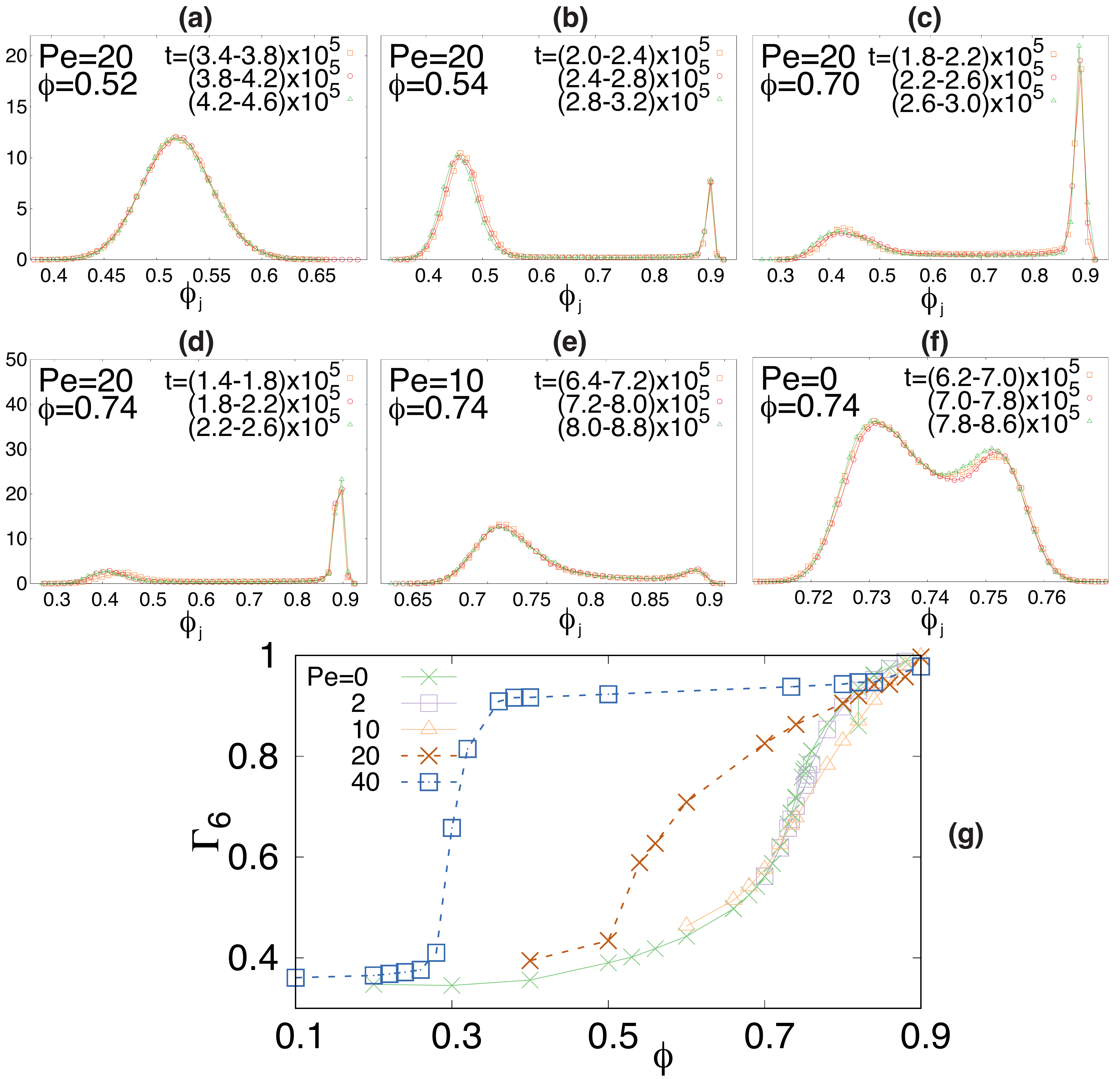}
	\vspace{-0.4cm}
	\caption{(Color online.) \textcolor{black}{(a-f)} Distributions of local 
	densities at $\phi = 0.52~\textcolor{black}{\textrm{(a)}}, \ 0.54~\textcolor{black}{\textrm{(b)}}, \ 0.7~\textcolor{black}{\textrm{(c)}}$ and fixed Pe$ = 20$ (first row, random initial configurations), and Pe $= 20~\textcolor{black}{\textrm{(d)}}, \ 10~\textcolor{black}{\textrm{(e)}}, \ 0~\textcolor{black}{\textrm{(f)}}$ and $\phi = 0.74$ (second row, starting from equilibrium configurations of the passive  system).
	The various curves correspond to data sampled over different time intervals given in the keys. 
	\textcolor{black}{(g)} $\Gamma_6$ as a function of $\phi$ for several Pe.}
\label{fig:cinco}
\end{figure}

We now switch on activity. We first focus on $\phi=0.734$, a density within the interval of co-existence in the passive limit. 
In Fig.~\ref{fig:cuatro} we display the local \textcolor{black}{hexatic} order parameter of three instantaneous configurations 
obtained from the evolution at Pe = 10.  The snapshots above are for an initial  
configuration with co-existence between a dense region with a rough horizontal form and a sparse region around it. 
Below are the snapshots for an initial stationary state at Pe = 40 where the system is strongly segregated.
In the first case  the system  breaks the horizontal dense region and it later recreates dense clusters of 
approximately round form. These clusters turn independently of one another and
have different (time-dependent) local \textcolor{black}{hexatic} order. 
Movie M1 illustrates the aforementioned dynamics at  $\phi=0.74$, with more details on the cluster formation.
In the second case, dumbbells are 
progressively evaporated from the large and dense clusters until less packed and smaller 
ones attain a stable size. Simultaneously, the regions in between the dense clusters reach the target density of the 
sparse phase. The subsequent dynamics are the same as for the steady state reached in the upper series of 
snapshots.
In the inset we show the time-dependence of $\Gamma_6$  and we verify that the 
 two runs reach the same asymptote.
Therefore, independently of the initial conditions, the dynamics at 
$\phi=0.734$ and Pe = 10 approach a stationary limit with co-existence. The same occurs 
for all Pe at this $\phi$, even the very small ones (supplementary information on the Pe = 2 case 
is given in \textcolor{black}{Fig.~S8}). 

Having established that a small amount of activity does not destroy co-existence, 
we determine how it affects its limits by sweeping the parameters $\phi$ and Pe.
 
With the data for the coarse-grained local density at different pairs $(\phi, \ \mbox{Pe})$ 
we built the density distributions of Fig.~\ref{fig:cinco}\textcolor{black}{(a-f)}. 
In the first row Pe = 20 and  $\phi = 0.52, \ 0.54, \ 0.7$ (\textcolor{black}{Fig.~S9} shows the dynamics at 
$\phi = 0.54$). \textcolor{black}{Figure~\ref{fig:cinco}(a)} presents a static symmetric distribution around
the global packing fraction that is in the liquid phase close to the boundary. \textcolor{black}{Figure~\ref{fig:cinco}(b)} shows the emergence of 
a second  peak at a higher density, $\phi_j\simeq 0.9$, while the weight on lower density has displaced to a 
lower  value of $\phi_j$. \textcolor{black}{Figure~\ref{fig:cinco}(c)} confirms the presence of the 
peak at $\phi_j \simeq 0.9$, the height of which has notably increased. Consequently, the weight on 
smaller local densities decreased and moved towards a slightly smaller value.
The appearance of the second peak is our criterium to draw the upper boundary of the 
homogeneous phase, \textcolor{black}{also complemented by the analysis of the structure factor in Fig.~S11}. 
At higher packing fractions the position of the second peak does not vary but its height increases
at the expense of the one of the first peak. The 
upper boundary of the phase segregated region is naturally determined by the disappearance 
of the low density peak (configurations below and above the upper co-existence boundary 
are shown in \textcolor{black}{Fig.~S10}).

The second row in Fig.~\ref{fig:cinco} displays the Pe-dependence of the local density plots at
$\phi= 0.74$, inside the co-existence interval at Pe = 0. This analysis confirms 
continuity between the steady states in the passive and active cases, \textcolor{black}{see also 
Fig.~S14.}
The position of the high density peak moves towards larger $\phi_j$ for increasing Pe, indicating 
that the dense regions compactify, and accordingly the 
loose regions get more void. This fact reveals that 
segregation is more effective at higher Pe. \textcolor{black}{Continuity upon increasing activity was also observed 
by analyzing the position of the density peaks moving along lines corresponding to equal proportions of disordered and \textcolor{black}{hexatic} regions in the system, 
see Figs.~S13 and S15.}

The \textcolor{black}{hexatic} order can also be used to analyze the phase diagram.
At Pe = 0 we used the steep increase of $\psi_6$ (around $\phi=0.734$, see Fig.~\ref{fig:tres})
to locate the boundary between liquid to phase separated phases
since 
this quantity 
does not fluctuate much around its sample average. At finite Pe, instead, clusters with rather different 
values of $\psi_6$ co-exist and it is more convenient to use $\Gamma_6$ 
to study this boundary. In Fig.~\ref{fig:cinco}\textcolor{black}{(g)}
we show $\Gamma_6$ as a function of $\phi$ for various Pe values. There is little dependence on Pe for, say,
Pe $\lesssim 10$, while for larger values the shoulder moves towards smaller densities, signaling that the 
phase boundary becomes one between gas (very low $\phi$) and segregated phases at higher Pe.

\textcolor{black}{
An analysis of the statistics of the bead displacements at different time-delays and 
 $(\phi,$ Pe) is given in Figs.~S16 and S17 where, in particular,
we distinguish the dynamics of  the liquid and segregated dumbbells, for more details see the SM.}
Movies M2-M4 complement this survey with emphasis on coarsening at Pe =~2,~10,~20.

Putting these results together we drew the phase diagram in Fig.~\ref{fig:phase-diagram}. The figure also 
includes some configurations 
at parameter values close to the limits of co-existence that clearly show liquid, phase separated and 
\textcolor{black}{hexatic} order. 
The lower boundary of the co-existence region decreases with increasing Pe since 
large activity favors the formation of high density clusters and therefore co-existence.
Furthermore, co-existence is allowed at higher 
global packing fractions. 
This is because  the
regions with \textcolor{black}{hexatic} order become denser and  leave more free space for 
the liquid phase under higher Pe~\footnote[3]{We did not consider here the transition between the \textcolor{black}{hexatic} and 
crystalline phases (dashed line in Fig.~\ref{fig:phase-diagram}) 
that has been studied, for a colloidal active system, in~\cite{Bialke12}. \textcolor{black}{The role of topological defects in the ordering of active crystalline phases has been studied in \cite{weber2014defect}}}. 
We conclude that we do not see any discontinuity between  the behaviour of the system 
at Pe = 0 and Pe $>$ 0 at the densities at which there is phase co-existence in the passive limit.

As in a conventional liquid-vapor transition, it is hard to establish where the first-order transition lies
with high precision.  It would be desirable to complement our analysis with a thermodynamic study of the 
phase transitions. The double transition scenario proposed in~\cite{BeKr11}
for the $2d$ passive hard disk problem was confirmed by the 
finite-size analysis of the equation of state, or packing-fraction dependence of the 
pressure in the NVT ensemble~\cite{BeKr11,Engel13}.  In contrast, the existence of an 
equation of state in generic active matter remains open. Indeed, 
the difficulty to precisely define a pressure with the properties of a state variable in active systems was 
underlined in a number of papers, see e.g.~\cite{Solon15,*solon2015pressure,Joyeux16,Patch16}. The results 
here presented should further stimulate the search for a consistent definition of 
pressure for (molecular) active matter and promote new studies of phase diagrams in other active systems.
 
\begin{acknowledgments}
Simulations ran on IBM Nextscale GALILEO at CINECA (Project INF16-fieldturb) 
under CINECA-INFN agreement and at Bari ReCaS e-Infrastructure funded by MIUR through  PON Research and
Competitiveness 2007-2013 Call 254 Action I.
GG acknowledges  MIUR for funding (PRIN 2012NNRKAF).
This research was supported in part by the National Science Foundation under 
Grant No. NSF PHY-1125915. LFC is a member of the Institut Universitaire de France. We thank R. Golestanian 
for early discussions. 
\end{acknowledgments}

\bibliography{dumbbells-biblio}

\end{document}